\documentclass[%
 reprint,
 amsmath,amssymb,
 aps,
]{revtex4-2}

\usepackage{graphicx}
\usepackage{dcolumn}
\usepackage{bm}
\usepackage{float}
\usepackage{ragged2e}

\begin{document}

\preprint{APS/123-QED}

\title{Open-Path Methane Sensing via Backscattered Light in a Nonlinear Interferometer}

\author{Jinghan Dong}
 \email{ip20409@bristol.ac.uk}
\author{Weijie Nie}%
\author{Arthur C. Cardoso}%
 \altaffiliation[Also at ]{Danish Fundamental Metrologi, Kogle Allé 5, 2970 Hørsholm, Denmark}
\author{Haichen Zhou}%
\author{Jingrui Zhang}%
\author{John G. Rarity}%
\author{Alex S. Clark}
\affiliation{ 
Quantum Engineering Technology Labs, H.H. Wills Physics Laboratory and School of Electrical, Electronic and Mechanical Engineering, University of Bristol, Bristol, UK}

\date{\today}

\begin{abstract}
Nonlinear interferometry has widespread applications in sensing, spectroscopy, and imaging. However, most implementations require highly reflective mirrors and precise optical alignment, drastically reducing their versatility and usability in outdoor applications. This work is based on stimulated parametric down conversion (ST-PDC), demonstrating methane absorption spectroscopy in the mid-infrared (MIR) region by detecting near-infrared (NIR) photons using a silicon-based CMOS camera. The MIR light, used to probe methane, is diffusely backscattered from a Lambertian surface, experiencing significant transmission loss. We implement a single-mode confocal illumination and collection scheme, using a two-lens system to mode-match the interfering beams to achieve background methane detection at a distance of 4.6 meters under a 60~dB loss. Our method is also extended to real-world surfaces, such as glass, brushed metal, and a leaf, showing robust background methane sensing with various target materials.\\
\end{abstract}

\maketitle


\section{Introduction}

Accurate methane detection is critical for monitoring emissions from natural gas infrastructure~\cite{zhang2020quantifying}, agriculture, and other anthropogenic sources, in order to mitigate climate change~\cite{masson2021climate} and to prevent explosion-related safety hazards caused by methane leaks in populated areas~\cite{kundu2016review}.

Optical gas sensors, if properly designed, can give us real-time measurement with high resistance to environmental disturbances such as humidity and low cross-sensitivity to other gases\cite{hodgkinson2012optical}, giving them advantages over conventional semiconductor and electrochemical sensors\cite{aldhafeeri2020review}. In particular, methane exhibits strong and highly specific absorption feature in the mid-infrared (MIR) region, which makes methane optical sensors unaffected by common atmospheric components like water vapor and carbon dioxide. The absorption in the MIR can exceed that in the 1.66~$\mu$m band by more than a factor of 60. However, most common optical methane sensors still rely on the 1.66~$\mu$m absorption spectrum, such as tunable diode laser absorption spectroscopy (TDLAS)~\cite{lackner2007tunable, liu2015highly} and cavity ring-down spectroscopy (CRDS)~\cite{fawcett2002trace}. This is because shortwave infrared (SWIR) detectors are more cost-effective, exhibit lower noise, typically do not need cooling systems, and are less affected by thermal background compared to MIR detectors.

Nonlinear interferometers based on spontaneous parametric down-conversion (SPDC) enable sensing and imaging with undetected photons by exploiting the principle of induced coherence~\cite{zou1991induced}. In such systems, photon pairs (signal and idler) are generated, and information in the idler can be obtained by detecting only the signal photons. This method has found broad application in quantum sensing~\cite{lindner2023high,tashima2024ultra}, imaging~\cite{lemos2014quantum,pearce2023practical,pearce2024single}, spectroscopy~\cite{hashimoto2024fourier}, microscopy~\cite{kviatkovsky2020microscopy}, tomography~\cite{vanselow2020frequency}, and holography. 

However, SPDC-based nonlinear interferometers have fundamental limitations. The coherence length of SPDC processes is typically short, restricting their suitability for long-range or open-path sensing and imaging. Moreover, these systems often rely on strong back-reflection of the idler photons using mirrors and require precise alignment. In realistic high-loss scenarios, such as targets with low reflectivity or high surface roughness (e.g. walls, white paper, or natural surfaces), the idler suffers significant losses, degrading the interference visibility and reducing the signal-to-noise ratio (SNR), so that the visibility and transmittance change is difficult to detect.

High-gain SPDC regimes~\cite{hashimoto2024fourier,machado2020optical} can address the low SNR issue, but typically require high-peak-power pulsed lasers and are still limited by coherence length. Furthermore, the spectral resolution of SPDC-based systems is inherently limited by the broadband signal and idler spectra. Upconversion-based methods have also been employed for open-path methane sensing, including demonstrations using mid-infrared (MIR) pulsed lasers and scattering targets with reflectivity of 0.2~\cite{Meng:18}, as well as intracavity enhancement schemes~\cite{Wolf:17}.

In our work, we build upon SPDC-based gas spectroscopy methods and introduce a seeding source~\cite{cardoso2024methane,dong2025methane} to implement stimulated parametric down-conversion (ST-PDC). Even with low seeding power, ST-PDC enables significantly enhanced signal intensity compared to spontaneous processes. We use a continuous-wave (CW) pump and a MIR seeding laser to generate NIR signal photons in a nonlinear crystal. While the MIR light probes methane, taking advantage of its strong and spectrally specific absorption features, the detection is performed at NIR wavelengths, where low-cost and low-noise silicon detectors can be used.

The introduction of the seeding beam can significantly improve the system's SNR, allowing interference fringes and visibility change to be clearly observed even in high-loss and weak methane absorption conditions. Unlike SPDC-based schemes that require mirrors for idler reflection, our approach supports the usage of real-world backscattering targets. We demonstrate robust methane sensing over long distances using diffuse targets, enabled by spatial mode-matching between the two generated signal beams, an unbalanced interferometer configuration~\cite{florez2022enhanced,Gemmell2023}, and effective homodyne detection. By monitoring visibility changes in the interference pattern, we extract ultra-low methane concentrations in ambient air. Our method also supports methane detection using a variety of real-world materials as targets.

\section{Experimental setup}

\begin{figure}[t!]
    \centering   \includegraphics[width=\linewidth]{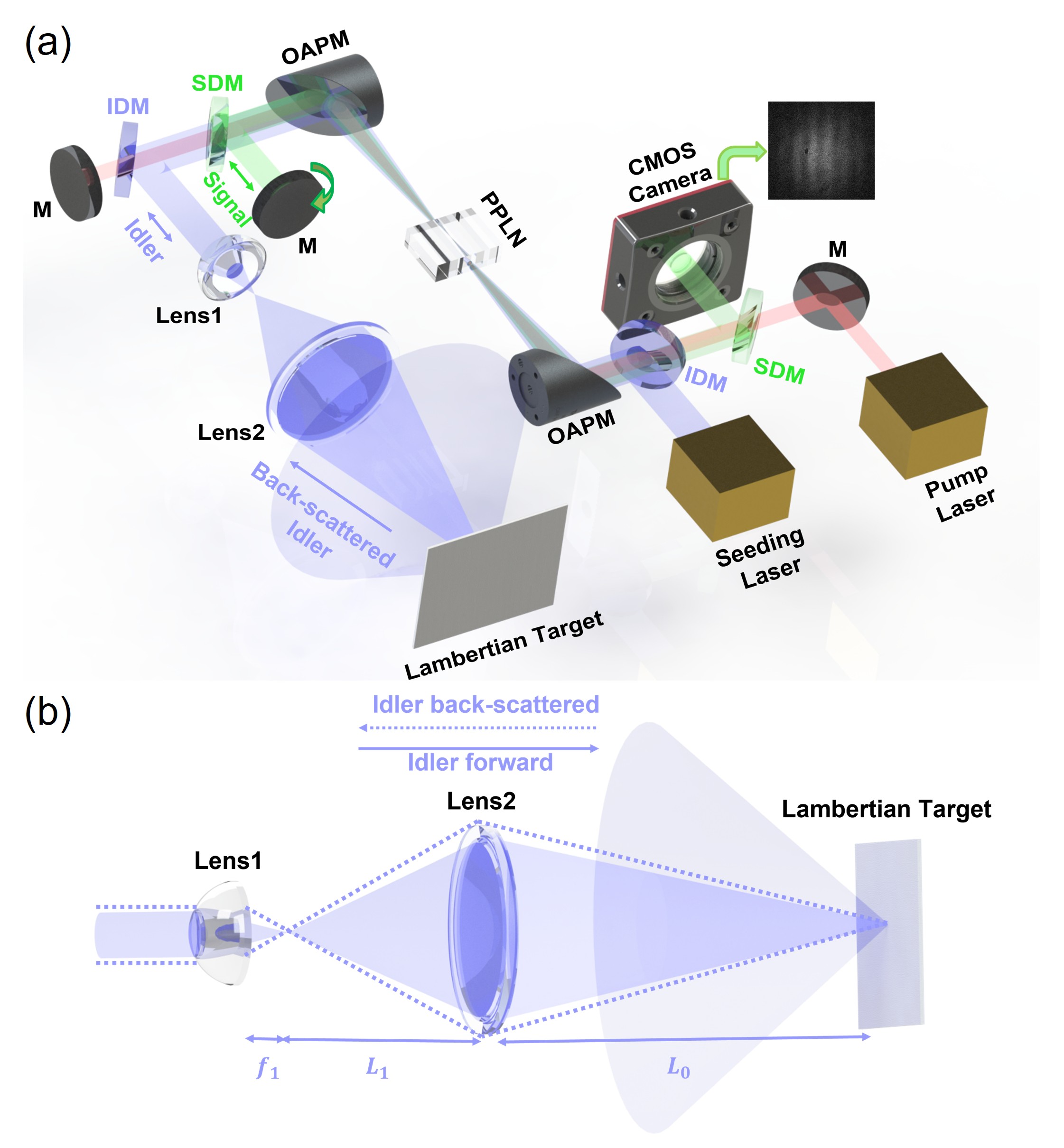}
    \caption{\justifying Schematic of the experimental setup. (a) A Michelson-type nonlinear interferometer is employed. The seeding idler laser and the backscattered idler, together with the pump, generate Signal~1 and Signal~2, respectively, via ST-PDC in a PPLN crystal. The spatial interference fringes produced by Signal~1 and Signal~2 can be obtained by the CMOS camera. M: mirror; SDM: signal dichroic mirror; IDM: idler dichroic mirror. (b) A two-lens system used for collecting backscattered idler. The dotted line indicates the optical path of the backscattered idler that couples back to the interferometer. \( f_1 \): focal length of Lens~1. \( L_1 \): distance between the focal point of Lens~1 and Lens~2. \( L_0 \): distance between Lens~2 and Lambertian target.}
    \label{fig:setup}
\end{figure}

Our experimental setup is illustrated in Fig.~\ref{fig:setup}(a). A 671~nm pump laser with an output power of 320~mW, along with a tunable seeding laser at 3.2~$\mu$m with a power of 0.6~mW, are both focused into a 10~mm long magnesium oxide-doped periodically poled lithium niobate (MgO:PPLN) crystal using an off-axis parabolic mirror (OAPM) with a focal length of 150~mm. The first 848~nm signal (Signal~1) is then generated via ST-PDC. Another identical OAPM collimates the pump, signal, and idler beams. The pump and Signal~1 are then reflected back into the crystal by separate mirrors, while the idler beam is scattered by a white paper target and stimulates a second PDC process in crystal, producing Signal~2. By slightly tilting the mirror that reflects Signal~1, vertical spatial fringes resulting from the interference between the angled Signal~1 and Signal~2 can be observed on a silicon-based CMOS camera (Thorlabs: CS165MU).

The two-lens system used to collect the backscattered idler beam shown in Fig.~\ref{fig:setup}(b) implements single-mode confocal illumination and collection, ensuring spatial mode matching and enhancing interference contrast. The collimated idler beam is first focused by Lens~1, which has a focal length \( f_1 = 18 \,\mathrm{mm} \) and a diameter of 12.7~mm. According to the lens equation
\begin{align}
\frac{1}{L_1} + \frac{1}{L_0} = \frac{1}{f_2} \text{,} \label{eq:lens} \tag{1}
\end{align}

\noindent where \( f_2 = 200 \,\mathrm{mm} \) is the focal length of Lens~2, for different distance \( L_0 \) to the Lambertian target, we can adjust \( L_1 \) to focus the idler beam onto the target. The idler beam is scattered by the Lambertian target, and its intensity distribution theoretically follows Lambert's cosine law: \( I = I_0 \cos \theta \)~\cite{pedrotti2017introduction}, where \( I_0 \) is the input intensity in the direction normal to the surface, and \( \theta \) is the angle between the surface normal and the observation direction.
The dashed line in the figure represents the optical path of the backscattered idler light that is collected and returned to the interferometer. Due to the nature of diffuse reflection, the light fills the entire 50~mm diameter aperture of Lens~2 regardless of the target distance \( L_0 \), and is then collimated by Lens~1, being focused back into the crystal by the OAPM.

By scanning the seeding laser over on- and off-resonance wavelengths, the interference visibility corresponding to each wavelength can be obtained. At on-resonance wavelengths, absorption of the idler by methane reduces the intensity of Signal~2, resulting in a lower visibility. The transmittance at each wavelength can be calculated from the visibility, defined as\cite{cardoso2024methane}
\begin{align}
V = \frac{2 \sqrt{\alpha} \tau}{1 + \alpha \tau^2}\text{,} \label{eq:visibility} \tag{2}
\end{align}
\noindent where \( \alpha = {I_{s_2}}/{I_{s_1}} \) is the intensity ratio of Signal~2 and Signal~1 measured at the camera at off-resonance wavelengths, related to the system loss, and \( \tau \) is the single-pass transmittance through the gas. 

By acquiring the entire absorption spectrum and fitting it with the HITRAN database~\cite{gordon2022hitran2020}, the background methane concentration can be obtained. A more detailed discussion of data processing and calculations is available in our previous work\cite{dong2025methane}.

\section{Signal mode-matching simulation}

Unlike specular reflection, Idler~2 returning to the crystal via diffuse reflection may not preserve the same spatial mode as Idler~1. This mismatch can result in a different spatial mode for Signal~2 compared to Signal~1. To address this, we performed beam propagation simulations to identify a lens configuration that ensures mode matching between Signal~1 and Signal~2. 

The initial collimated beam diameters of the pump and Idler~1 are measured as 1.20~mm and 2.66~mm, respectively. The beam diameter \( D \) of the backscattered idler after being collimated by Lens~1 and Lens~2 can be estimated using the geometrical relation \( D = d_2 \cdot (f_1 / L_1) \), where \( d_2 = 50\,\mathrm{mm} \) is the diameter of Lens~2. Combined with Eq.~\eqref{eq:lens}, this indicates that \( D \) varies with the target distance \( L_0 \). 

\begin{figure}[t!]
    \centering   \includegraphics[width=0.8\linewidth]{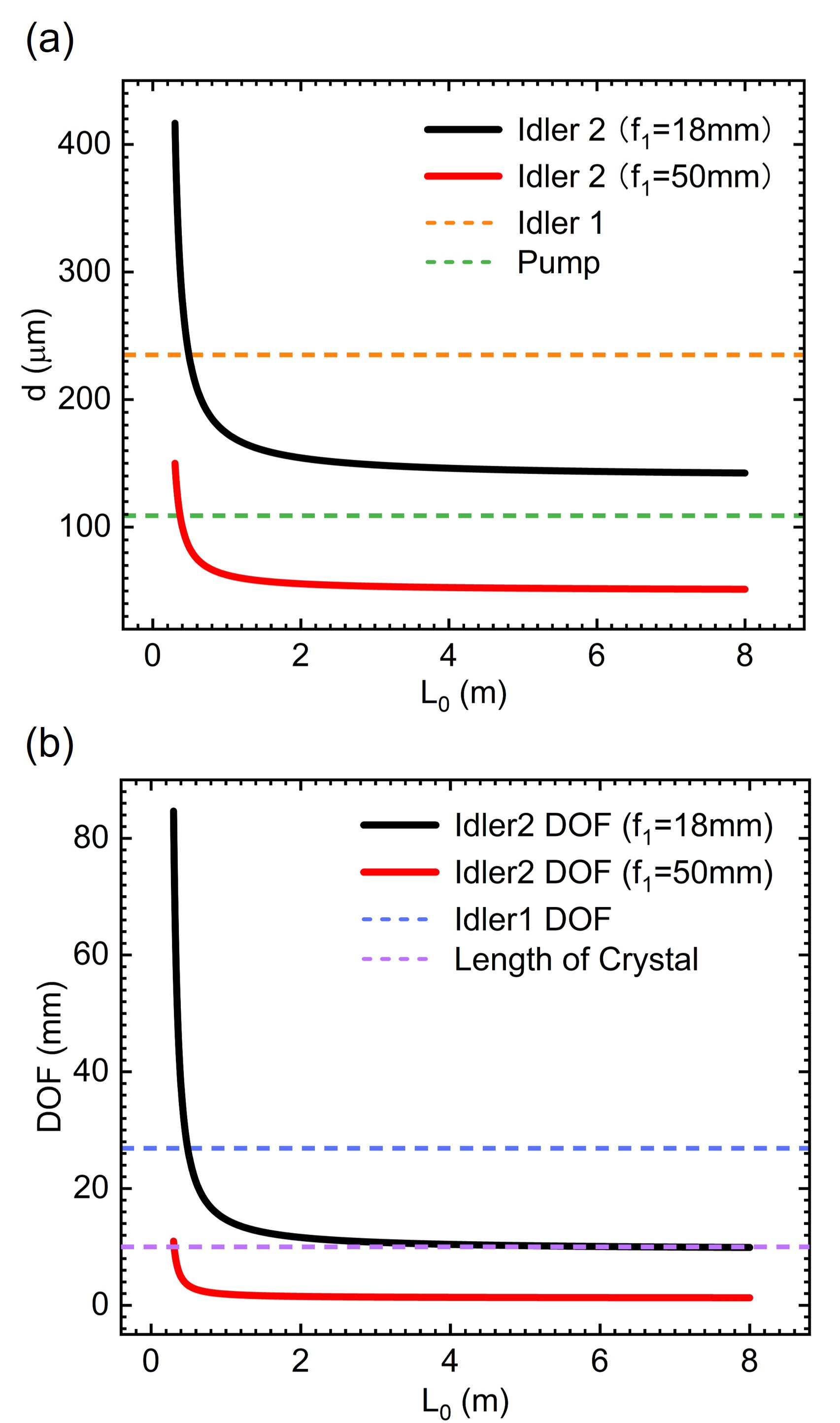}
    \caption{\justifying Simulation results. The black and red solid lines represent the (a) beam diameter and (b) depth of focus of Idler~2 at the crystal when using Lens~1 with focal lengths of \( f_1 = 18\,\mathrm{mm} \) and \( f_1 = 50\,\mathrm{mm} \), respectively, as the distance \( L_0 \) between Lens~2 and the Lambertian target is varied. In (a), the orange and green dashed lines represent the beam diameters of Idler~1 and the pump in the crystal, respectively. In (b), the blue and purple dashed lines indicate the depth of focus of Idler~1 in the crystal and the length of the crystal, respectively.}
    \label{fig:simulation}
\end{figure}

In Fig.~\ref{fig:simulation}(a), the green and orange dashed lines represent the focused beam diameters of the pump and Idler~1 in the crystal, which are 109~$\mu$m and 235~$\mu$m, respectively. These are calculated using the Gaussian beam waist formula \( d = \frac{4 \lambda f}{\pi D} \)~\cite{siegman1986lasers}, where \( \lambda \) is the wavelength, \( f \) is the focal length of the OAPM, and \( D \) is the beam diameter at the OAPM. 

Since the pump beam diameter is smaller than that of Idler~1, the beam diameter of Signal~1 is determined by the pump. The solid black line in Fig.~\ref{fig:simulation}(a) shows the simulated beam diameter of Idler~2 at the crystal for different \( L_0 \) values, using our final choice of Lens~1 with \( f_1 = 18\,\mathrm{mm} \). Although the beam diameter of Idler~2 decreases rapidly with increasing \( L_0 \), becoming smaller than that of Idler~1, it remains larger than the pump beam diameter. As a result, the generated Signal~2 maintains a beam size the same as Signal~1. Moreover, as \( L_0 \) increases, the beam diameter of Idler~2 at the crystal becomes smaller than that of Idler~1, which helps enhance the intensity of Idler~2 within the nonlinear crystal and thus increases the intensity of Signal~2. The solid red line in Fig.~\ref{fig:simulation}(a) represents the Idler~2 beam diameter at the crystal when using Lens~1 with \( f_1 = 50\,\mathrm{mm} \). It can be seen that an improper choice of focal length for Lens~1 leads to a rapid decrease in the Idler~2 beam diameter relative to the pump as \( L_0 \) increases. As a result, the generated Signal~2 becomes much smaller than Signal~1, and the resulting mode mismatch between the two signals reduces the interference visibility and undermines the reliability of the fitted visibility values.

\begin{figure}[t]
    \centering   \includegraphics[width=\linewidth]{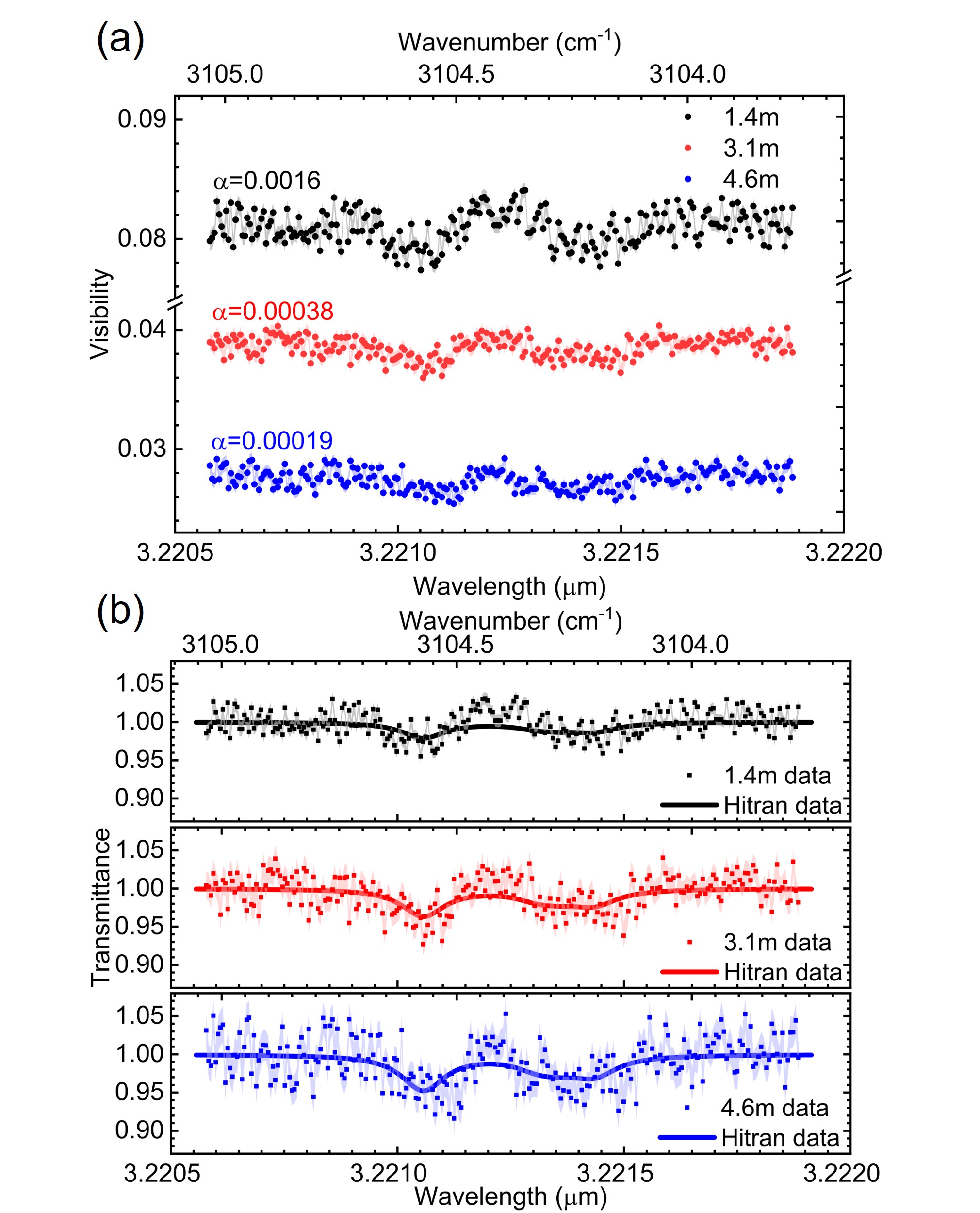}
    \caption{\justifying (a) The visibility change over on- and off-resonance wavelengths, and (b) the absorption spectrum of background methane at different detection distances.}
    \label{fig:diff_dis}
\end{figure}

While maintaining the same beam size for the two signals generated, it is also essential to ensure that Idler~2 remains approximately collimated in the crystal. This requires that the depth of focus (DOF), defined as twice the Rayleigh range \( \mathrm{DOF} = 2z_R = \frac{2\pi w_0^2}{\lambda} \), exceeds the crystal length. In Fig.~\ref{fig:simulation}(b), the purple and blue dashed lines indicate the crystal length (10~mm) and the DOF of Idler~1 (26.9~mm) after being focused by the OAPM, respectively. The solid black line shows the simulated DOF of the backscattered Idler~2 at the crystal for different target distances \( L_0 \) with \( f_1 = 18\,\mathrm{mm} \). Although the DOF of Idler~2 decreases with increasing \( L_0 \) and becomes lower than the DOF of Idler 1, it remains longer than or comparable to the crystal length. This ensures that the second ST-PDC process remains collinear, similar to the first, thereby maximizing the ST-PDC efficiency. The solid red line in Fig.~\ref{fig:simulation}(b) shows the depth of focus (DOF) of Idler~2 when an incorrect focal length of \( f_1 = 50\,\mathrm{mm} \) is used for Lens~1. As \( L_0 \) increases, the DOF of Idler~2 rapidly drops below the crystal length (10~mm), reaching as low as approximately 1.2~mm. This implies that only a small portion of the crystal supports effective collinear generation of Signal~2.

Since our system operates in the stimulated parametric down-conversion (ST-PDC) regime with a fixed signal wavelength generation, a shortened DOF leads to the presence of non-phase-matched idler momentum components (\( \vec{k} \)), reducing conversion efficiency. Additionally, the increased divergence of Idler~2 may cause the beam to exceed the PPLN crystal aperture.
Even in the case of spontaneous PDC, such divergence can reduce the indistinguishability of the photon pairs, resulting in lower interference visibility. Both focal lengths -- \( f_1 = 50\,\mathrm{mm} \) and our final choice of \( f_1 = 18\,\mathrm{mm} \) -- were tested experimentally, and the results are discussed in the next section.

\section{Experimental results}

\subsection{\label{sec:dist}Background methane detection with Lambertian target}
Using white paper as the target, we reduced the intensity of Signal~1 by a factor of 100. By adjusting \( L_1 \), the distance \( L_0 \) to the target can be varied. At distances of \( L_0 = \) 1.4~m, 3.1~m, and 4.6~m, corresponding to round-trip absorption paths of 2.8~m, 6.2~m, and 9.2~m, we tuned the idler wavelength and measured the resulting interference visibility, as shown in Fig.~\ref{fig:diff_dis}(a). At each distance, two visibility dips were observed, corresponding to the absorption of the idler by methane at on-resonance wavelengths. Each data point represents the average of 600 measurements, with an exposure time of 13~ms per frame on the CMOS camera. The off-resonance visibility values at 1.4~m, 3.1~m, and 4.6~m were 8.10\%, 3.91\%, and 2.79\%, respectively. The corresponding \(\alpha\) values are 0.0016, 0.00038, and 0.00019. At a distance of 4.6~m, the SNR of the visibility measurement is $\sim40$.

The solid angle subtended by the lens from the perspective of the scattering target is given by \( \Omega = A_{\text{lens}} / L_0^2 \), where \( A_{\text{lens}} \) is the area of Lens~2. According to Lambert’s cosine law, the total collected power is \( P_0 = \int_\Omega I_0 \cos(\theta) \, d\Omega \). Under a small-angle, near-axis approximation (\( \cos \theta \approx 1 \)), we obtain \( P_0 \propto 1 / L_0^2 \). This theoretical scaling is confirmed by our measurements. The \(\alpha\) values exhibit a \(1/L_0^2\) dependence, indicating a quadratic increase in system loss with distance. In contrast, the observed visibility decreases more gradually, approximately following a \(1/L_0\) trend.

\begin{figure*}[t!]
    \centering   \includegraphics[width=\linewidth]{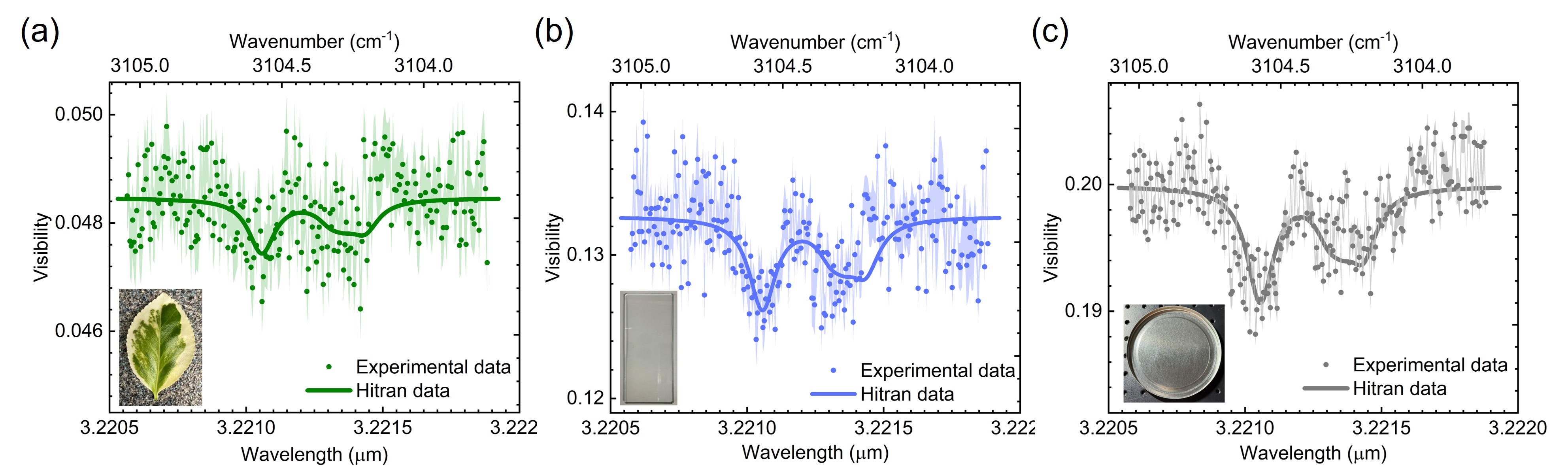}
    \caption{ Visibility variation at on- and off-resonance wavelengths for (a) a leaf target at 1.4~m, (b) a glass target at 4.6~m, and (c) a brushed metal plate target at 4.6~m, compared with the corresponding HITRAN absorption data.}
    \label{fig:diff_material}
\end{figure*}

The experimental data presented here was taken with a lens pair with focal lengths of 18~mm and 200~mm. However, it is worth noting that we initially used lenses of 50~mm and 200~mm focal lengths. In doing so we found that due to mode mismatch between Signal~1 and Signal~2, the system exhibited a loss that increased much more rapidly with target distance \( L_0 \) than the expected quadratic scaling (\( \sim 1/L_0^2 \)) predicted by geometric and Lambertian considerations. As a result, although a starting visibility of 24.80\% was obtained when \( L_0 \) is 30~cm, the interference fringes were hard to observe when \( L_0 \) reached only 1~m. This experimental result is consistent with the mode-matching simulation results in Fig.~\ref{fig:simulation} and further underscores the critical importance of achieving proper spatial mode matching between the interfering signal beams.

Furthermore, for the optimized lens configuration (\( f_1 = 18\,\mathrm{mm} \)) we attempted to directly detect the backscattered 3.2~µm idler beam using a MIR sensor, but were unable to measure any power. This highlights the advantage of our approach, which leverages homodyne detection and the use of a more sensitive silicon-based detector. Also, the ``single mode'' ST-PDC phase-matching acts as a wavevector filter, which eliminates substantial MIR thermal background noise~\cite{ma2023thermal}.

Using Eq.~\eqref{eq:visibility}, we computed the transmittance spectrum at each distance with error bands, as shown in Fig.~\ref{fig:diff_dis}(b). Solid lines are theoretical absorption spectra from the HITRAN database~\cite{gordon2022hitran2020}. By fitting the measured data to the HITRAN data, we were able to retrieve the background methane concentrations at each distance: \( (2.96 \pm 0.29) \)~ppm, \( (3.10 \pm 0.32) \)~ppm, and \( (2.89 \pm 0.34) \)~ppm for path lengths of 1.4~m, 3.1~m, and 4.6~m, respectively.

\subsection{\label{sec:targets}Background methane detection with real-world targets}

To demonstrate the broader applicability of our method in real-world scenarios, we performed measurements using a variety of common materials as reflective targets. Signal~1 was attenuated by 100 times for all measurements. 

Figure~\ref{fig:diff_material}(a) shows the measured methane absorption spectrum using a leaf as the target, at a distance of 1.4~m. The green data points with error bands represent the experimental visibility as a function of the idler wavelength. Using the same transmittance calculation method described above, we fit the spectrum to the HITRAN data and obtained an ambient methane concentration of \( (2.96 \pm 0.36) \)~ppm. The solid green line represents the theoretical visibility curve, calculated using an \(\alpha\) value of 0.00059 derived from the measured off-resonance visibility of 4.85\% and the HITRAN-based transmittance data. Compared to the white paper target result at the same distance, the leaf exhibits a lower reflectance, indicated by the lower off-resonance visibility in comparison to the white paper.

Figures~\ref{fig:diff_material}(b) and~\ref{fig:diff_material}(c) present the visibility spectra for glass and a brushed metal plate, respectively, both measured at a distance of 4.6~m. The blue and gray round data points with error bands correspond to the experimental results, with off-resonance visibility values of 13.26\% and 19.97\%, resulting in \(\alpha\) values of 0.0044 and 0.01, respectively. The corresponding solid lines are the theoretical visibilities calculated from the HITRAN transmittance data. The measured methane concentrations using the glass and metal plate targets are \( (3.02 \pm 0.27) \)~ppm and \( (2.85 \pm 0.22) \)~ppm, respectively.

\subsection{Specular and diffuse reflection robustness}
Compared to using a specular reflective target, employing a diffuse reflection target is more representative of real-world applications. As the reflected intensity follows a $\cos\theta$ angular distribution, slight deviations of the incident beam angle from the surface normal do not significantly reduce the efficiency of Idler~2 coupling back into the nonlinear interferometer.
In principle, this removes the need for highly precise alignment. At long distances, mirrors are very sensitive to angular alignment, making angle-dependent measurements challenging. Moreover, under high-visibility conditions, it becomes easier to observe visibility variations. Therefore, to compare angular sensitivity, we evaluated the system using both a mirror and a Lambertian target placed 5~cm from the collection lens.

To clearly observe interference fringes, we reduced the intensity of Signal~1 by a factor of 100. We first measured the maximum visibility of 53.7\% when the idler was at normal incidence to the Lambertian target. To ensure a fair comparison, the intensity of Idler~2 was reduced when using the mirror such that the visibility was close to that of the Lambertian case, yielding 55.3\%.

\begin{figure}[t]
    \centering   \includegraphics[width=0.9\linewidth]{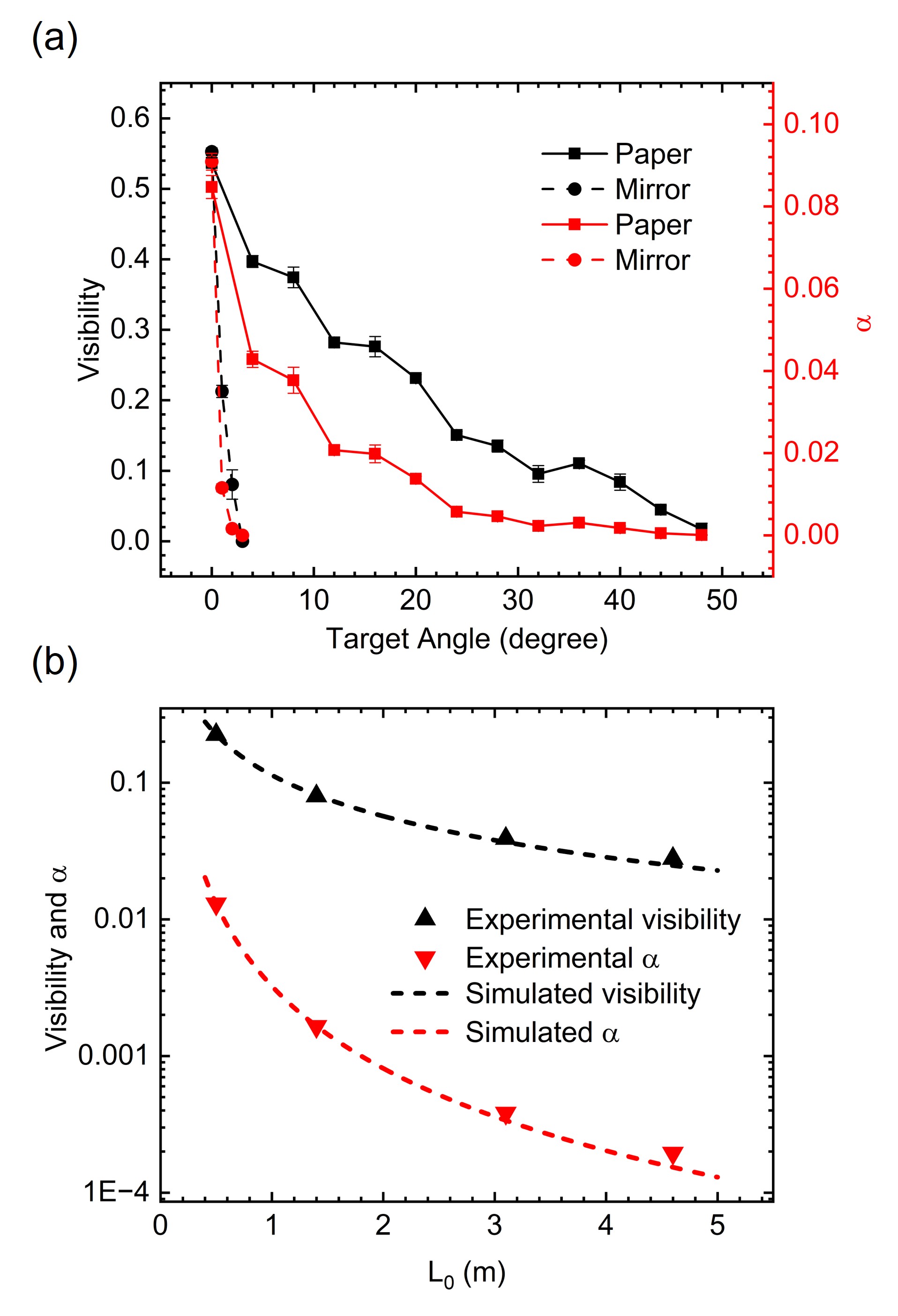}
    \caption{\justifying (a) Experimental results of interference visibility and $\alpha$ as functions of the target rotation angle, using either paper or a mirror as the reflecting surface. (b) Simulation and experimental results of visibility and $\alpha$ at different distances to a white paper target.
}
    \label{fig:a_V}
\end{figure}

As shown in Fig.~\ref{fig:a_V}(a), when the target was rotated, the visibility in the mirror case dropped rapidly, becoming close to zero at a rotation angle of $3^\circ$. In contrast, for the white paper (Lambertian) target the visibility decreased gradually with angle, remaining detectable until $48^\circ$, beyond which it approached zero. This demonstrates that our system is more robust when applied to real-world surfaces, as opposed to ideal laboratory reflectors such as mirrors.

The black and red triangular markers in Fig.~\ref{fig:a_V}(b) show the off-resonance visibility measured experimentally and the corresponding \(\alpha\) at different distances using white paper as the scattering target. In addition to the data discussed in Section~\ref{sec:dist}, we also measured a visibility of 22.5\% at \(L_0 = 0.5\)~m. However, due to insufficient accumulation of methane, no absorption spectrum was observed at 0.5~m. In this logarithmic scale plot, it is evident that the visibility decreases more slowly with distance than the loss-related \(\alpha\).

According to the analysis in Section~\ref{sec:dist}, the idler power collected by the lens is proportional to \(1/L_0^2\). Therefore, the loss-related parameter \(\alpha\) is also expected to scale with \(1/L_0^2\). Based on the visibilities measured at different distances, we extracted the corresponding \(\alpha\) values and fitted them with \(\alpha = \alpha_0/L_0^2\), where \(\alpha_0\) is a constant related to the intrinsic system loss. Following Eq.~(\ref{eq:visibility}), the visibility data can be subsequently fitted using the function $\frac{2\sqrt{\alpha_0 / L_0^2}}{1 + \alpha_0 / L_0^2}$. When \(\alpha_0\) is small, as is the case for a diffuse scattering target, then this simplifies to $2\sqrt{\alpha_0} / L_0$ and we can see that visibility scales as \(1/L_0\) with distance. The black and red dashed lines in Fig.~\ref{fig:a_V} (b) represent the fitted curves for visibility and \(\alpha\), respectively. In both Fig.~\ref{fig:a_V}(a) and (b), we observe that the visibility does not decrease linearly with increasing system loss. Instead, visibility decreases more slowly than \(\alpha\), indicating our system is robust under high-loss conditions.

\section{Conclusion and discussion}

This work demonstrates, for the first time to our knowledge, background gas sensing using undetected backscattered MIR light. We present the application of a nonlinear interferometer to detect background methane using a diffuse reflection target under high-loss conditions. We propose and validate the importance of selecting the focal lengths in a two-lens collection system to efficiently collect backscattered MIR probe light in nonlinear interferometry. The mode matching between Signal~2, which is generated from backscattered idler photons, and Signal~1 is shown to be critical for preserving interference visibility.

We measured background methane using a white paper target at various distances. The experimental observation that visibility scales as \(1/L_0\), while the system loss increases quadratically, highlights the advantage of our homodyne detection scheme. Our results demonstrate a minimum detectable methane concentration of 8.3~ppm·m, and a minimum detectable methane concentration change of approximately 300~ppb.

In our previous work~\cite{dong2025methane}, detection at a distance of 11~m was achieved using high-reflectivity mirrors. The maximum detection distance using a Lambertian target is currently limited to 4.6~m in this work, as the visibility becomes too low to be detected at longer ranges. As previously reported, it is possible to balance the interferometer by adjusting the pump polarization for the first and second ST-PDC processes, thereby equalizing the intensities of Signal~1 and Signal~2 and improving visibility. This approach can enhance sensitivity by increasing the visibility change. However, in our current setup, Signal~1 has already been attenuated by a factor of 100, and Signal 2 is ultra-low due to the scattering target. Further attenuation leads to an unacceptably low signal-to-background ratio due to the increased contribution from SPDC, which cannot be effectively filtered out. This is because the second-generation signal, seeded by the backscattered idler, becomes too weak to dominate over the SPDC background. As a result, even if a narrowband filter is applied for detection, the signal photons generated from SPDC, which are spectrally indistinguishable from the seeded signal, remain a significant source of background. If higher seeding power can be used, further reduction of Signal~1 could improve the interference visibility and enable longer-distance methane detection.

We also demonstrated measurements using a variety of common real-world materials, including a leaf, brushed metal and glass, which highlights the practical applicability and robustness of our approach. Furthermore, this approach can be extended to the detection of other gases and potentially to imaging based on backscattered light.

\bibliography{apssamp}

\end{document}